\documentclass[12pt,english,pre,unsortedaddress,nofootinbib]{revtex4-1}
\usepackage[T1]{fontenc}
\usepackage[latin9]{inputenc}
\usepackage{xcolor}
\usepackage{pdfcolmk}
\usepackage{amsmath}
\usepackage{amssymb}
\usepackage{graphicx}
\usepackage{wasysym}
\PassOptionsToPackage{normalem}{ulem}
\usepackage{ulem}

\makeatletter

\providecommand{\tabularnewline}{\\}
\providecolor{lyxadded}{rgb}{0,0,1}
\providecolor{lyxdeleted}{rgb}{1,0,0}
\DeclareRobustCommand{\lyxsout}[1]{\ifx\\#1\else\sout{#1}\fi}

\usepackage{float}
\usepackage{array}
\usepackage{color}
\usepackage{subcaption}
\usepackage{ragged2e}
\DeclareCaptionJustification{justified}{\justifying}
\usepackage{enumitem}
\usepackage{setspace}

\floatstyle{ruled}
\newfloat{algorithm}{H}{loa}
\providecommand{\algorithmname}{Algorithm}
\floatname{algorithm}{\protect\algorithmname}

\usepackage{babel}

\makeatother

\usepackage{babel}
\begin{document}
\global\long\def\V#1{\boldsymbol{#1}}%
 
\global\long\def\M#1{\boldsymbol{#1}}%
 
\global\long\def\Set#1{\mathbb{#1}}%

\global\long\def\D#1{\Delta#1}%
 
\global\long\def\d#1{\delta#1}%

\global\long\def\norm#1{\left\Vert #1\right\Vert }%
 
\global\long\def\abs#1{\left|#1\right|}%

\global\long\def\grad{\M{\nabla}}%
 
\global\long\def\avv#1{\langle#1\rangle}%
 
\global\long\def\av#1{\left\langle #1\right\rangle }%

\global\long\def\P{\mathcal{P}}%

\global\long\def\ki{k}%
 
\global\long\def\wi{\omega}%

\global\long\def\slip{\breve{\V u}}%

\global\long\def\bu{\V u}%
 
\global\long\def\bv{\V v}%
 
\global\long\def\br{\V r}%

\global\long\def\sM#1{\M{\mathcal{#1}}}%
 
\global\long\def\fM#1{\M{\mathfrak{#1}}}%
 
\global\long\def\Mob{\sM M}%
 
\global\long\def\J{\sM J}%
 
\global\long\def\S{\sM S}%
 
\global\long\def\L{\sM L}%
 
\global\long\def\R{\sM R}%

\global\long\def\N{\sM N}%
 
\global\long\def\K{\sM K}%
 
\global\long\def\slipN{\breve{\N}}%

\global\long\def\slipW{\breve{\V W}}%
 
\global\long\def\rot{\M{\Psi}}%
 
\global\long\def\Rot{\M{\Xi}}%

\global\long\def\eqd{\overset{d}{=}}%

\global\long\def\Donev#1{{\bf [{\color{red}#1}]}}%
 
\global\long\def\bren#1{{\bf [{\color{blue}#1}]}}%
 
\global\long\def\bcom#1{{\bf [{\color{magenta}#1}]}}%

\title{Sedimentation of a Colloidal Monolayer Down an Inclined Plane}
\author{B. Sprinkle, S. Wilken, S. Karapetyan, M. Tanaka, Z. Chen, J. R. Cruise,
B. Delmotte, M. M. Driscoll, P. Chaikin, and A. Donev}
\begin{abstract}
We study the driven collective dynamics of a colloidal monolayer sedimentating
down an inclined plane. The action of the gravity force parallel to
the bottom wall creates a flow around each colloid, and the hydrodynamic
interactions among the colloids accelerate the sedimentation as the
local density increases. This leads to the creation of a universal
``triangular'' inhomogeneous density profile, with a traveling density
shock at the leading front moving in the downhill direction. Unlike
density shocks in a colloidal monolayer driven by applied torques
rather than forces {[}Phys. Rev. Fluids, 2(9):092301, 2017{]}, the
density front during sedimentation remains stable over long periods
of time even though it develops a roughness on the order of tens of
particle diameters. Through experimental measurements and particle-based
computer simulations, we find that the Burgers equation can model
the density profile along the sedimentation direction as a function
of time remarkably well, with a modest improvement if the nonlinear
conservation law accounts for the sub-linear dependence of the collective
sedimentation velocity on density.
\end{abstract}
\maketitle

\section{Introduction}

The dynamics of active and driven colloidal suspensions is interesting
not only because of its inherent out-of-equilibrium nature, but also
because it is often dominated by collective effects leading to the
formation of large-scale structures and flows. In the Stokes (overdamped)
limit relevant to colloids, the hydrodynamic interactions between
the particles are long-ranged and strongly depend on the presence
of nearby boundaries such as confining walls. While self-propelled
colloids (microswimmers) are of great interest, externally-driven
colloids present a simpler system to analyze and study both analytically
and via computer simulations. In particular the only many-body and
long-ranged interactions in driven suspensions are hydrodynamic interactions
created by the generated solvent flow. While confining walls generally
screen hydrodynamic interactions to be less long-ranged than in bulk
suspensions, the nature of the hydrodynamic flows generated by the
activity or external driving mechanism crucially affects the resulting
collective behavior.

In prior work \cite{Rollers_NaturePhys,MagneticRollers,TwoLines_Rollers,NonlocalShocks_Rollers,RollersLubrication},
some of us studied the collective behavior of microrollers: magnetic
colloids sedimented above a bottom floor (wall) in an external magnetic
field rotating around an axis parallel to the wall. Each spinning
particle propels itself parallel to the wall, but the collective motion
in a non-dilute suspension is much faster than that of an isolated
colloid. For microrollers, the driving mechanism is an applied torque,
and the flow field created by a single particle corresponds to a rotlet
above a no-slip wall \cite{blake1971note,OseenBlake_FMM}. This flow
advects other particles both in the vertical direction (away or toward
the wall), as well as in the transverse directions parallel to the
wall. The resulting collective dynamics is surprisingly rich even
at moderate densities, for which in the absence of the external drive
the colloids would form a single monolayer with in-plane packing fractions
$\phi\apprle0.5$. Uniform microroller suspensions develop a two-layer
structure with a slow bottom layer and a fast top layer \cite{RollersLubrication},
and non-uniform suspensions develop traveling-wave density shocks
\cite{NonlocalShocks_Rollers} that are unstable to transverse perturbations
\cite{TwoLines_Rollers}, leading to a fingering instability that
can create stable motile clusters of colloids (critters) held together
entirely by hydrodynamic interactions \cite{Rollers_NaturePhys}.

For the case of a non-uniform suspension of microrollers, the formation
and dynamics of a density front can be described by a \emph{non-local}
conservation law, and the front has a finite width that is proportional
to the typical height of the particles above the wall \cite{NonlocalShocks_Rollers}.
Here we study a similar system of colloids sedimented above a bottom
wall but now apply a force, rather than a torque, to drive collective
dynamics. This can easily be accomplished in the lab simply by tilting
the bottom wall at an angle $\theta$ and letting the colloids sediment
down the inclined plane. The flow field created by a single particle
corresponds to a Stokeslet above a no-slip wall \cite{blake1971note,OseenBlake_FMM},
and advects other particles primarily in the direction of the motion.
As we demonstrate, the resulting collective density dynamics can be
modeled rather accurately by a \emph{local} conservation law that
can be approximated by a Burgers equation. This leads to sharp density
fronts in the form of propagating shock solutions of the inviscid
Burgers equation. Similar Burgers-like shocks have been observed for
driven and active suspensions confined in a narrow slit channel (top
and bottom walls) \cite{BurgersDroplets_Bartolo,Beatus2009,Tsang2016,Lefauve2014},
but a crucial difference is that the local flow field around a particle
in these cases is quasi-two-dimensional and corresponds to the flow
created by a potential dipole. While colloidal diffusion in a flat
monolayer \cite{DiffusionColloidalMonolayer} and sedimentation in
a vertical channel \cite{ColloidalRT_Gompper} have been studied both
using simulations and experiments, to our knowledge the sedimentation
of colloids down an inclined plane has not been studied before. Carpen
and Brady studied theoretically the sedimentation of a colloidal suspension
in a thicker slit channel and predicted a transverse instability created
by unstable stratification due to shear-induced particle migration
\cite{Carpen2002}. Here we study a monolayer sedimenting above a
single wall and find different dynamics that we study by a combination
of experiments, computer simulations, and theory.

We begin by discussing the problem theoretically at a mean-field continuum
level in Section \ref{sec:Burgers}, and then present results from
experiments and particle-based simulations and compare them to the
predictions of the theory in Section \ref{sec:Results}.

\section{\label{sec:Burgers}Burgers Model}

\begin{figure}
\centering{}\includegraphics[width=1\textwidth]{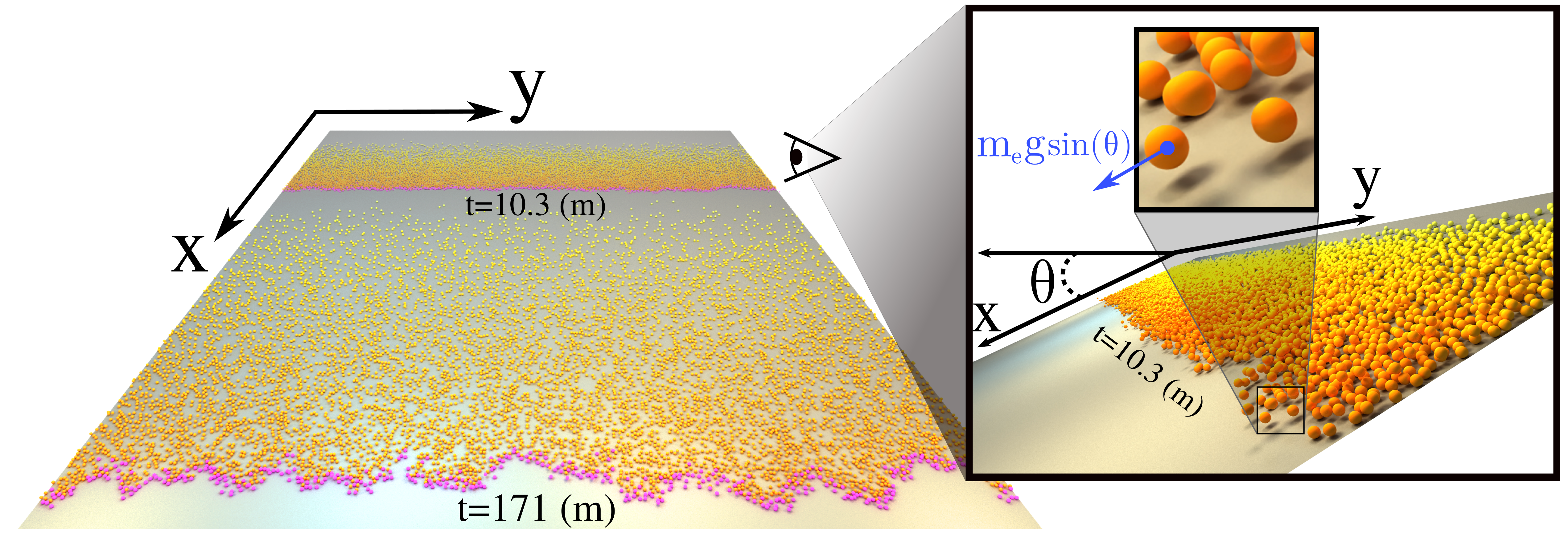}
\caption{\label{fig:simulations}The left panel shows a frontal view of two
snapshots from a typical simulation of a particle suspension sedimenting
down an inclined plane. We show the particles at two separate times:
the first (top) at a time $t=10.3$min shortly after the particles
have begun to sediment down the plane, and the second at $t=171$min
when the triangular density profile is well formed (see text). The
front of the suspension is highlighted by magenta particles, visually
emphasizing the coarsening of the density front in time (see Section
\ref{subsec:Transverse}). The right panel shows a side view of the
suspension at $t=10.3$min along with an inset zoom showing several
particles near the front at their typical Brownian height $h_{g}$.
An animated version of the left panel is available in the Supplementary
Information.}
\end{figure}

We consider a monolayer of spherical colloids of radius $a$ sedimented
above a bottom wall tilted at angle $\theta$, as illustrated in Fig.
\ref{fig:simulations}. We take the $x$ axis to be in the direction
of the gravitational acceleration $g\sin\theta$, and the $z$ axis
to be in the direction perpendicular to the wall and pointing away
from the gravitational pull $\sim g\cos\theta$. Here we consider
the quasi-one-dimensional problem where initially the colloids are
uniformly distributed in the unbounded (in the theory, bounded at
millimeter scales in the experiments) or periodic (in simulations)
$y$ direction, and in the $x$ direction the colloids are initially
contained in the \emph{finite} interval $[x_{b}(t=0),\;x_{f}(t=0)].$
We wish to quantify the collective dynamics as the colloids sediment
downhill as time $t\geq0$ elapses. Specifically, we will average
along the $z$ direction and consider the coarse-grained \emph{number}
density of colloids $\rho(x,y,t)$ in the $xy$ plane (parallel to
the bottom wall). Although estimates of the Péclet number $\text{Pe}=aV_{\mathrm{avg}}\left(k_{B}T/\left(6\pi\eta a\right)\right)^{-1}\sim10$
(where $V_{avg}\in[0.1,0.3]$ is estimated based on Fig. \ref{fig:v_phi})
are not high, we will neglect in-plane diffusion in our continuum
models. Also, we will for now assume there is no transverse instability
and thus the density remains quasi-one-dimensional, $\rho(x,y,t)\approx\rho(x,t)$,
even though the left panel in Fig. \ref{fig:simulations} shows that
this is not exactly true at later times. We discuss the validity of
this assumption in more detail in Section \ref{subsec:Transverse},
and conclude that the roughness of the front remains small compared
to the extent of the density profile in the $x$ direction for all
times.

Let us assume that an unbounded uniform suspension with $\rho(x,t)\approx\rho_{0}$
sediments down the plane (in the positive $x$ direction) with collective
velocity $v(\phi_{0})$, where $\phi=\rho\pi a^{2}$ is the in-plane
packing fraction and $a$ is the radius of the colloids. We will use
$\phi(x,t)$ instead of $\rho(x,t)$ as a more physically-intuitive
variable. The collective sedimentation velocity can be estimated as
a function of packing fraction experimentally or via computer simulations,
as we discuss in Section \ref{subsec:Uniform}. Because particles
interact hydrodynamically and advect other particles via the flows
they create, the function $v(\phi)$ is quite generally monotonically
increasing. For small $\phi\ll1$ we expect the linear approximation
\begin{equation}
v\left(\phi\right)\approx v_{0}+\frac{S}{2}\phi,\label{eq:v_phi_Burgers}
\end{equation}
to hold, where $v_{0}$ is the sedimentation velocity of an isolated
particle, and $S/2$ is a constant that measures the strength of the
hydrodynamic interactions.

It is natural to expect that the density dynamics follows the \emph{local}
conservation law
\begin{equation}
\frac{\partial\phi}{\partial t}+\frac{\partial}{\partial x}\left(v\left(\phi\right)\phi\right)=0,\label{eq:subBurgers}
\end{equation}
which at low densities is expected to be well-approximated by the
Burgers equation
\begin{equation}
\frac{\partial\phi}{\partial t}+v_{0}\frac{\partial\phi}{\partial x}+S\phi\frac{\partial\phi}{\partial x}=0.\label{eq:Burgers}
\end{equation}
It is important to note that for colloidal microrollers, the local
mean-field approximation \eqref{eq:subBurgers} is not appropriate,
and one must take into account the nonlocal nature of hydrodynamic
interactions \cite{NonlocalShocks_Rollers}. However, for sedimentation,
the hydrodynamic interactions are more local and we expect that the
local approximation \eqref{eq:subBurgers} is suitable; we come back
to this question in Section \ref{sec:Conclusions}.

Adopting the Burgers approximation \eqref{eq:subBurgers} we can obtain
several analytical results. We will set $v_{0}=0$ for convenience
since a nonzero self velocity simply translates the solution to the
right by $v_{0}t$. Let us first consider an initial density profile
that is a square wave, $\phi(x,t=0)=\phi_{0}$ for $0\leq x\leq L$,
and zero otherwise. The Burgers equation \eqref{eq:Burgers} can be
solved analytically, showing that at the front of the square wave
there is a propagating shock with position $x_{f}(t)$, with $x_{f}(0)=L$,
and at the back of the square wave, which stays fixed at $x_{b}(t)=0$,
there is a rarefaction wave. The rarefaction and shock waves meet
at time
\begin{equation}
t_{\text{triang}}=2\frac{L}{S\phi_{0}},\quad\text{when}\quad x_{f}\left(t_{\text{triang}}\right)=2L.\label{eq:t_triang}
\end{equation}
After this initial transient, there is a \emph{triangular density
profile}
\begin{equation}
\phi\left(x,t\geq t_{\text{triang}}\right)=\begin{cases}
\frac{x}{x_{f}(t)}\sqrt{\frac{2M}{St}}=\frac{x}{St} & \text{if}\quad0\leq x\leq x_{f}(t)\\
0 & \text{otherwise}
\end{cases},\label{eq:triangle_wave}
\end{equation}
where the total conserved ``mass'' is 
\begin{equation}
M=\int\phi(x,t)dx=L\phi_{0},\label{eq:mass}
\end{equation}
and the shock wave at the front of the triangle is located at
\begin{equation}
x_{f}(t)=\sqrt{2SMt}.\label{eq:x_f_Burgers}
\end{equation}

This remarkably simple triangle-wave solution is, in fact, universal
and independent of the initial conditions. In particular, it was proven
by Lax that any compactly-supported initial condition will for long
times asymptotically approach the triangle wave solution \eqref{eq:triangle_wave}
\cite{HyperbolicLaws_Lax}. It was further shown by Goodman that shock
solutions of the Burgers equation in two dimensions are stable against
perturbations in the transverse (i.e., the $y$) direction \cite{Burgers2D_ShockStability}.
The reader should note that while the equations of Stokes flow are
time-reversible, once shocks form the Burgers equation is no longer
time reversible\footnote{For example, many initial conditions can reach the same triangle solution
\eqref{eq:triangle_wave} after some finite time \cite{HyperbolicLaws_Lax}.} because of the implicit dissipation at the front caused by the neglected
diffusion.

\section{\label{sec:Results}Results: Simulations and Experiments}

We now compare results from experimental measurements and computer
simulations to the predictions of the Burgers model. We begin by describing
briefly the experimental setup and simulation methodology, and then
present results for the collective sedimentation velocity $v\left(\phi\right)$
for uniform suspensions, before studying the formation and propagation
of Burgers density shocks. For the results presented here we incline
the bottom floor at an angle $\theta=45^{\circ}$, though we studied
other angles as well and found that the Péclet number controls the
layering of the suspension. For larger angles ($\theta\gtrsim60^{\circ}$),
at larger packing densities some of the particles get lifted away
from the floor and the colloids are not in a monolayer, while for
smaller angles the sedimentation velocity is low and diffusion becomes
important. For $\theta=45^{\circ}$ and the Péclet numbers considered
in this work we find that the suspension remains in a monolayer.

\subsection{\label{subsec:Experiments}Experiments}

We performed sedimentation experiments with colloidal monolayers sedimenting
down an inclined plane at angle $\theta=45^{\circ}\pm3^{\circ}$ placed
inside of a custom-built tilting microscope that sets the angle. The
colloids are polystyrene spheres (Duke Scientific) of diameter $d=2a=\left(4.2\pm0.1\right)\,\mu$m
suspended in mixtures of water ($H_{2}O$) and heavy water ($D_{2}O$)
to modify the gravitational height $h_{g}=k_{B}T/\left(m_{e}g\right)$,
where $g$ is the acceleration of gravity, $m_{e}$ is the buoyant
mass, and the temperature $T$ is $\left(22\pm3\right){}^{\circ}$C.
Note that the normal distance between the surface of the particles
and the bottom wall is $\sim k_{B}T/\left(m_{e}g\cos\theta\right)$.
We use three mixtures with different mass fraction of heavy water:
0\% heavy water ($h_{g}=0.250\mu m$), 12.5\% heavy water ($h_{g}=0.336\mu m$),
and 25\% heavy water ($h_{g}=0.500\mu m$). The solution also contains
50mM BIS-TRIS to stabilize the particles.

The sample chamber is a rectangular borosilicate capillary tube with
dimensions $6\times25\times0.3$mm. The glass surface is coated with
polyelectrolyte multilayers, specifically, three layers each of PDADMAC(+)
and PSS(-) to prevent particle sticking to the glass surface. The
colloidal suspensions are loaded into capillaries at a three-dimensional
volume fraction $5.0\times10^{-4}$ and the sample chamber is centrifuged
first along the $z$-axis to rapidly drive particles to the bottom
surface, and then centrifuged along the $x$-axis ``uphill\textquotedbl{}
to create a densely-packed, partially crystallized monolayer with
area fraction of $0.7\pm0.1$, extending over a distance of approximately
100 particle diameters at the bottom of the capillary. This is the
initial condition of all sedimentation instability experiments. Experiments
to measure sedimentation velocity as a function of density for uniform
suspensions are prepared the same way without the $x$-axis centrifugation.

\begin{figure}
\centering{}\includegraphics[width=1\textwidth]{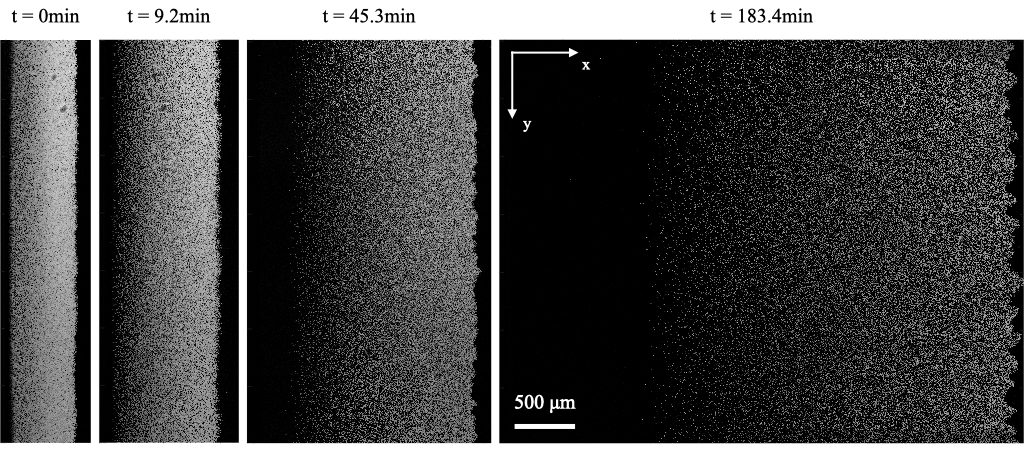}
\caption{\label{fig:experiment_profile}Microscope images of the density profile
for experiment $\#2$ at $h_{g}=0.336\ \mu\text{m}$ at several times.
The scattered light intensity is roughly proportional to the density.
As time progresses an inhomogeneous (triangular) density profile develops
(higher density on the right, the direction of motion of the front).
See Fig. \ref{fig:Exp_Burg_Fit} for averages along the $y$ axis,
and the middle panel in Fig. \ref{fig:front_rough} for images of
the front extracted from this data. An animated version of this figure
is available in the Supplementary Information.}
\end{figure}

Sedimentation density profiles are imaged with a custom built tilting
microscope. The camera, objective, sample, and back lighting (530nm
LED) are all mounted together and rotated about a single axis so that
prepared monolayers can be imaged while rotated from flat $\theta=0$
(relative to gravity) to an incline at angle $\theta$, while the
whole monolayer remains in the focal plane. A photograph of the microscope
can be found in the Supplementary Information. Sample snapshots from
the evolution of the monolayer density observed in experiments are
shown in Fig. \ref{fig:experiment_profile}. For experiments with
a uniform suspension we employ conventional particle tracking techniques
\cite{crocker1996methods} to count particles and obtain an accurate
estimate of $\phi$.

Large fields of view were needed to accurately measure the density
evolution over long periods of time, limiting the accuracy of particle
tracking techniques, especially at the highest packing fractions.
We therefore we use scattered intensity as a proxy for particle density
in the $xy$ plane, see Fig. \ref{fig:experiment_profile}. This is
not sufficiently accurate to obtain precise measurements of $\phi\left(x,y,t\right)$
since the intensity is not strictly linear in the density, and also
because of imaging artifacts; this is evidenced in the fact that the
total intensity is not conserved and fluctuates by $\sim10-20$\%
especially when the packing fraction is larger than $\phi\gtrsim0.4$
(see top of Fig. \ref{fig:Exp_Burg_Fit}). Near the end of the experiments,
when the density is lower, $\phi\lesssim0.2-0.3$, we are able to
locate particles reliably, as confirmed by the fact that the total
number of particles is conserved to a few percent. For such low densities
we confirm that the scattered light intensity is a good proxy for
packing fraction. The total number of particles is approximately conserved
among all Burgers shock experiments since we start with the same amount
of stock colloidal solution in each capillary and then add water and
$D_{2}O$ to change $h_{g}$. The total number of particles $N_{\text{\text{exp}}}\approx78500$
in the viewing area was estimated using particle tracking at later
times, with the length of the frame being $L_{y}=3344\mu$m, giving
the total (conserved) ``mass'' \eqref{eq:mass} to be $M_{\text{\text{exp}}}=N_{\text{\text{exp}}}\pi a^{2}/L_{y}\approx325\ \mu\text{m}$.
Knowing the total number of particles (total mass) allows us to rescale
the scattered intensity to a packing density, which can then be compared
to theoretical and computational predictions.

\subsection{\label{subsec:Simulations}Simulations}

The numerical methods used in our simulations are described in detail
in recent work by some of us \cite{RollersLubrication} on the collective
dynamics in uniform suspensions of microrollers. Briefly, our lubrication-corrected
Brownian Dynamics method simulates the translational and rotational
dynamics of all particles in the suspension in three dimensions. The
method is based on the method of Stokesian Dynamics and includes both
far-field hydrodynamics (at the Rotne-Prager level \cite{StokesianDynamics_Wall}
without stresslets) as well as semi-analytical lubrication corrections
due to particles coming close to other particles or the bottom wall.
We consistently and efficiently account for Brownian motion, which
is essential in the particle simulations in order to set the gravitational
height of the particles.

In the simulations we take the radius of the particles $a=2.1\mu$m
and the density of the particles is set to $\rho_{p}=1041\text{kg/m}^{3}$,
and we fix the temperature at $22^{\circ}$C. We compute the density
of the solvent $\rho_{f}$ from the concentration of $D_{2}O$ from
standard tables \cite{swift1939densities}; the values are indicated
in Table \ref{tab:sim_parameters}. Adding $D_{2}O$ to water also
changes the viscosity of the solution $\eta$ \cite{jones1936viscosity},
as indicated in Table \ref{tab:sim_parameters}. The steric repulsion
potential between the particles and between the particles and the
wall is approximated by \cite{RollersLubrication} 
\begin{equation}
\Phi(r)=\Phi_{0}\begin{cases}
1+\frac{d-r}{2a\delta_{\text{cut}}/\ln(10)} & r<d\\
\text{exp}\left(\frac{d-r}{2a\delta_{\text{cut}}/\ln(10)}\right) & r\geq d
\end{cases},
\end{equation}
where $r$ is the distance from a particle to another particle or
to the wall, $d=a(1-\delta_{\text{cut}})$ for particle-wall repulsion
and $d=2a(1-\delta_{\text{cut}})$ for particle-particle repulsion
\footnote{Note that while the case $r<d$ should not happen for hard particles,
slight overlaps do occur in numerical simulations.}. We set the strength of the repulsion to $\Phi_{0}=4k_{B}T$ and
take $\delta_{\text{cut}}=10^{-2}$ to mimic approximately hard sphere
interactions through a ``firm potential'' (see section 3c of \cite{RollersLubrication}).
To accurately resolve the suspension dynamics we set the time step
size to $\D t=4.88\times10^{-3}(6\pi\eta a^{3})/\left(k_{B}T\right)=0.2$s.
Given this comparatively small $\Delta t$, on the order of hundreds
of thousands of time steps are needed to simulate the long timescales
observed in our sedimentation experiments, which limits the number
of particles we can include in our simulations relative to the experiments.

\begin{table}
\centering{}%
\begin{tabular}{|c|c|c|c|}
\hline 
$D2O$\% & 0 & 12.5 & 25\tabularnewline
\hline 
$h_{g}=$ & $0.25\mu\text{m}$ & $0.336\mu\text{m}$ & $0.5\mu\text{m}$\tabularnewline
\hline 
$\rho_{f}=$ & $998\text{kg/m}^{3}$ & $1009\text{kg/m}^{3}$ & $1020\text{kg/m}^{3}$\tabularnewline
\hline 
$\eta=$ & $0.95$mPas & $0.98$mPas & $1.01$mPas\tabularnewline
\hline 
\hline 
$L_{y}=$ & $116\mu\text{m}$ & $106\mu\text{m}$ & $77\mu\text{m}$\tabularnewline
\hline 
$N=$ & $2759$ & $2592$ & $1811$\tabularnewline
\hline 
$T_{s}=$ & $5.3$min. & $10.2$min. & $38$min.\tabularnewline
\hline 
\end{tabular}\caption{\label{tab:sim_parameters}Parameters (rows) used in the simulations
for different gravitational heights (columns) $h_{g}=k_{B}T/m_{e}g$
of the colloidal monolayer.}
\end{table}

Our simulations are unbounded in the positive $z$ direction, which
is a good approximation to the experiments where $L_{z}/a\approx150\gg1$.
For uniform suspensions, we use a domain that is periodic with period
$L_{xy}$ in both the $x$ and $y$ directions; we controlled the
packing fraction by varying $L_{xy}$ while keeping the number of
particles fixed. For Burgers shocks the simulation domain is infinite
in the $x$ direction (the direction of motion) and periodic in the
$y$ (the spanwise) direction with period $L_{y}$ indicated in Table
\ref{tab:sim_parameters}; the smallest $L_{y}/a\sim37$, which we
have confirmed is sufficiently large to make periodic artifacts in
the hydrodynamics negligible.

To generate an initial configuration of particles at time $t_{0}$
confined to a region of the $xy$ plane of area $L_{x}\times L_{y}$,
we perform Markov Chain Monte Carlo equilibrium runs at a constant
in-plane packing fraction $\phi_{0}$ with $N_{0}$ particles. For
uniform suspensions, $\phi_{0}=\phi$ and we keep all $N=N_{0}=1024$
particles. For Burgers shock simulations, we need to generate an inhomogeneous
density profile. Therefore, we first generate a uniform one at the
maximum density and then randomly remove particles to reduce the density
where needed. Specifically, we take the initial density profile extracted
from experiments $\phi_{\text{exp}}\left(x,t_{0}\right),$ set \footnote{The factor of $1.1$ gives us a small buffer above the maximum.}
$\phi_{0}=1.1\max\phi_{\text{exp}}\left(x,t_{0}\right)$ and $N_{0}=4000$,
and independently and uniformly randomly remove each particle with
probability $\left(1-\phi_{\text{exp}}(x,t_{0})/\phi_{0}\right)$.
The final number of particles $N$ after this rejection step is indicated
in Table \ref{tab:sim_parameters}.

\subsection{\label{subsec:Uniform}Uniform suspensions}

\begin{figure}
\centering{}\includegraphics[width=0.75\textwidth]{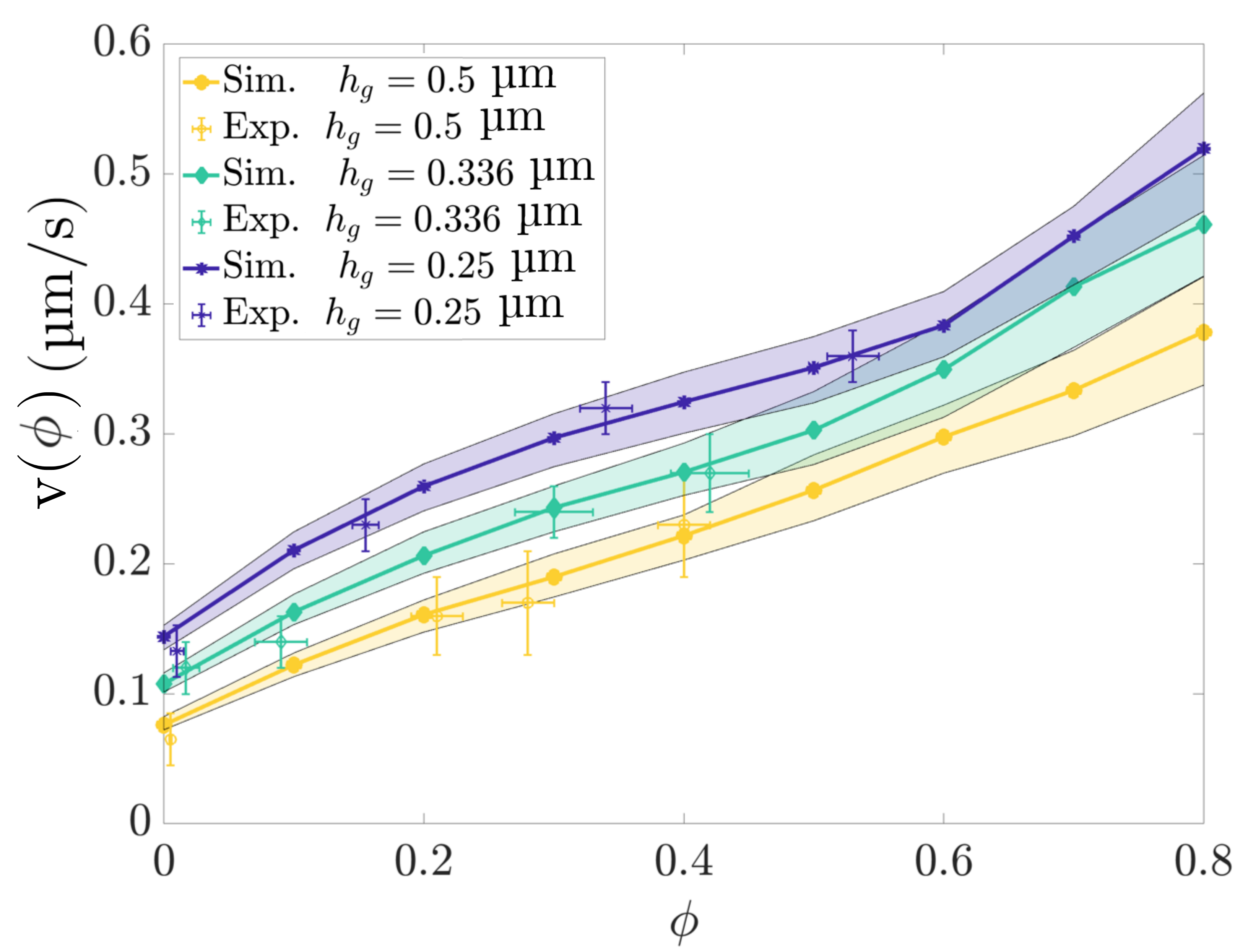}
\caption{\label{fig:v_phi}Collective sedimentation velocity as a function
of in-plane packing fraction for each of the three gravitational heights,
for both simulations and experiments. Experimental measurements of
$v(\phi)$ are shown as markers with error bars of one standard deviation.
The solid curve shows data from simulations at $\theta=45^{\circ}$;
the boundaries of the shaded region around the simulation data are
from simulations with $\theta=42^{\circ}$ (smaller velocity) and
$\theta=48^{\circ}$ (larger velocity), in order to account for the
experimental uncertainty in the sedimentation angle $\theta=45^{\circ}\pm3^{\circ}$.
In the simulations the colloids remained in a monolayer only for $\phi\lessapprox0.6.$}
\end{figure}

Our computational and experimental results for the collective sedimentation
velocity $v\left(\phi\right)$ at the three different gravitational
heights are shown in Fig. \ref{fig:v_phi}. The apparent $x$ velocity
of the particles is computed over intervals of $1s$ (but the results
are not sensitive to this choice), using particle tracking in the
experiments, and the average velocity is computed for each packing
fraction. We see a good agreement between the measured and predicted
collective velocities, to within experimental uncertainty. The relationship
is roughly linear as in \eqref{eq:v_phi_Burgers}. To obtain a more
accurate functional form of $v\left(\phi\right)$ to use in \eqref{eq:subBurgers},
we fit the simulation data with a rational function (typically with
degrees 3/2 but no larger than 5/2, not shown). Since the simulation
results for $v(\phi)$ shown in Fig. \ref{fig:v_phi} are sub-linear
over the range of densities of interest ($\phi<0.6$), we refer to
\eqref{eq:subBurgers} as the \textit{sub-Burgers equation}.

\subsection{\label{subsec:Shocks}Burgers shock waves}

In the remainder of this section we focus our attention to comparing
our experimental and simulation results to the prediction of the mean-field
sub-Burgers model \eqref{eq:subBurgers} and its Burgers approximation
\eqref{eq:Burgers}.

\subsubsection{\label{subsec:Transverse}Transverse (in)stability}

Before we discuss results for the one-dimensional density profiles
$\phi(x,t)$ modeled by Eqs. \eqref{eq:subBurgers} and \eqref{eq:Burgers},
we must understand whether the two-dimensional density profiles $\phi(x,y,t)$
are actually effectively one-dimensional, that is, whether the density
front remains stable to transverse perturbations.

\begin{figure}
\centering{}\includegraphics[width=1\textwidth]{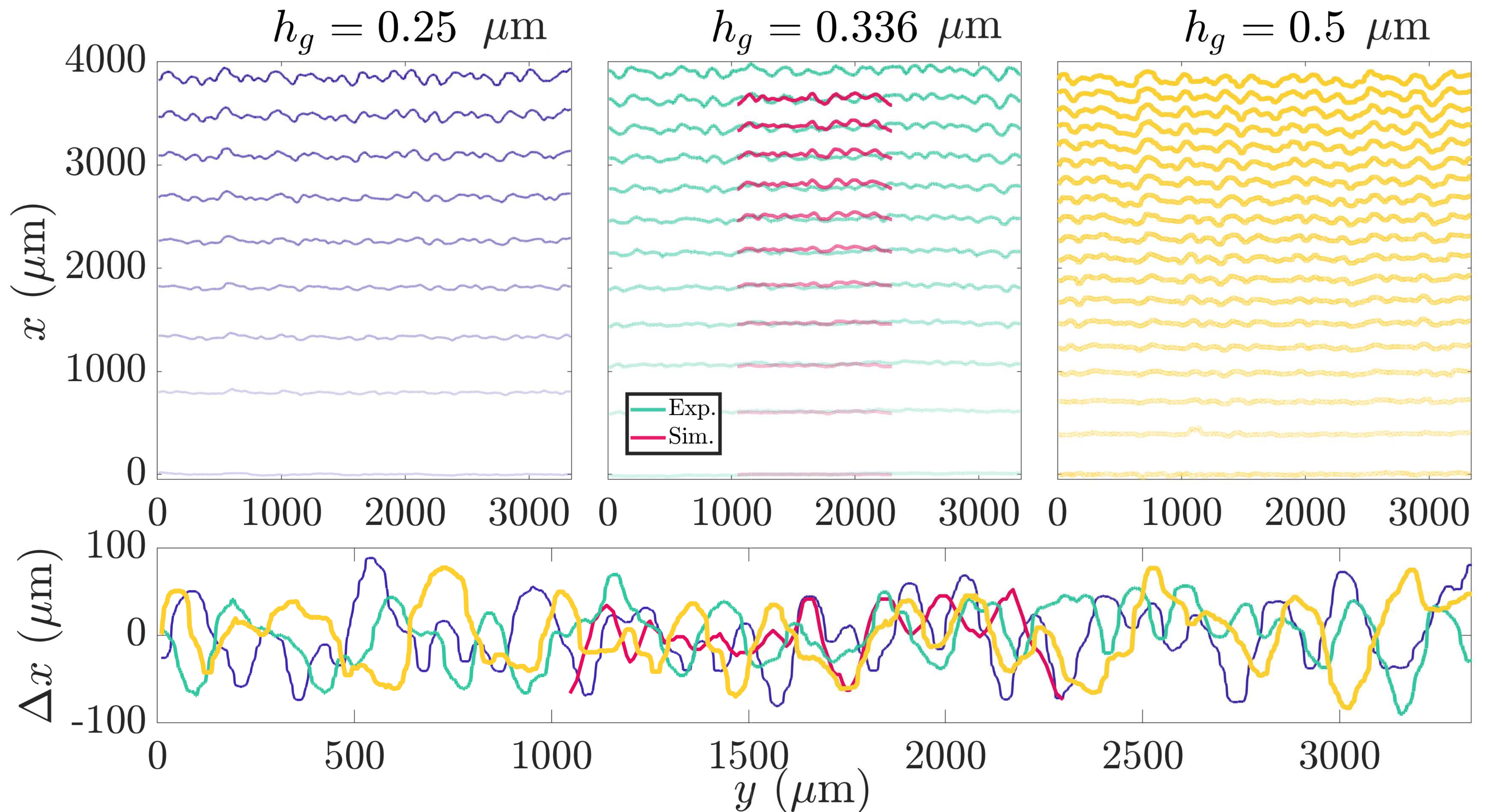}
\caption{\label{fig:front_rough}Position of the density front $x_{f}\left(y;t\right)$
in one of the experiments at each gravitational height. In the top
panels, we overlay snapshots of the density front in time increments
of 1000s, where the opacity of the curves increases with time. The
middle panel also shows the results of a particle simulation we ran
for $h_{g}=0.336$ using $N=8000$ particles with initial density
$\phi_{0}=0.6$, where the time is scaled by a factor of $2.6$ to
account for the different total number of particles and domain geometry.
The bottom panel overlays the shifted fronts $\protect\D x(y;t)=x_{f}\left(y;t\right)-\protect\av{x_{f}\left(y;t\right)}$
at the final time for each gravitational height, as well as the simulation
results for $h_{g}=0.336$. We see that the scale of the roughness
and characteristic ``wavelength'' of the front are roughly independent
of $h_{g}$ and are comparable in experiments and simulations.}
\end{figure}

To do this, from the experimental images and binned simulation densities,
we extract the position of the density front $x_{f}\left(y;t\right)$
such that the density is essentially zero for $x>x_{f}(y)$; see the
left panel of Fig. \ref{fig:simulations} for an illustration of the
particles determined to be at the density front at two different times.
The extracted fronts $x_{f}\left(y;t\right)$ are shown in Fig. \ref{fig:front_rough}.
For a rough comparison, we also perform a simulation at $h_{g}=0.336\mu\text{m}$
with $N=8000$ particles and a wider domain with periodic length $L_{y}=1260\mu$m,
and a uniform initial packing density of $\phi(0\leq x\leq L_{x},y,t)=0.6$
where $L_{x}=L_{y}/8.6$. We show the simulated density front profiles
overlayed on top of the experimental ones in the middle panel of Fig.
\ref{fig:front_rough}, but it should be noted that a direct comparison
is not possible because at present we cannot simulate a system of
the same dimensions as the experiments due to the very large number
of particles in the experiments. Due to the unmatched conditions between
the simulation and experiment, the shift in the mean position of the
front in the simulations has been scaled by an empirical factor of
$2.6$ to align the fronts with the experiments for easier visual
comparison. This factor is consistent with a rough estimate based
on the Burgers time scale \eqref{eq:t_triang} assuming the initial
condition in experiments is a square wave, $t_{\text{triang}}^{\text{(exp)}}/t_{\text{triang}}^{\text{(sim)}}=\left(L^{\text{(exp)}}\phi_{0}^{\text{(sim)}}\right)/\left(L^{\text{(sim)}}\phi_{0}^{\text{(exp)}}\right)\approx\left(500\mu\text{m}\cdot0.6\right)/\left(147\mu\text{m}\cdot0.65\right)\approx3.1$.

The results in Fig. \ref{fig:front_rough} show that the density front
becomes rough over time, however, they do not show the formation of
a transverse instability at a precise wavelength, as is the case for
microrollers \cite{MagneticRollers,TwoLines_Rollers}. Instead, the
roughness has a broad range of characteristic length scales (as determined
from the correlation function of $x_{f}\left(y;t\right)$, not shown)
and pin-pointing a precise ``wavelength'' of the roughness is not
possible due to the large statistical uncertainty.

\begin{figure}
\centering{}\includegraphics[width=1\textwidth]{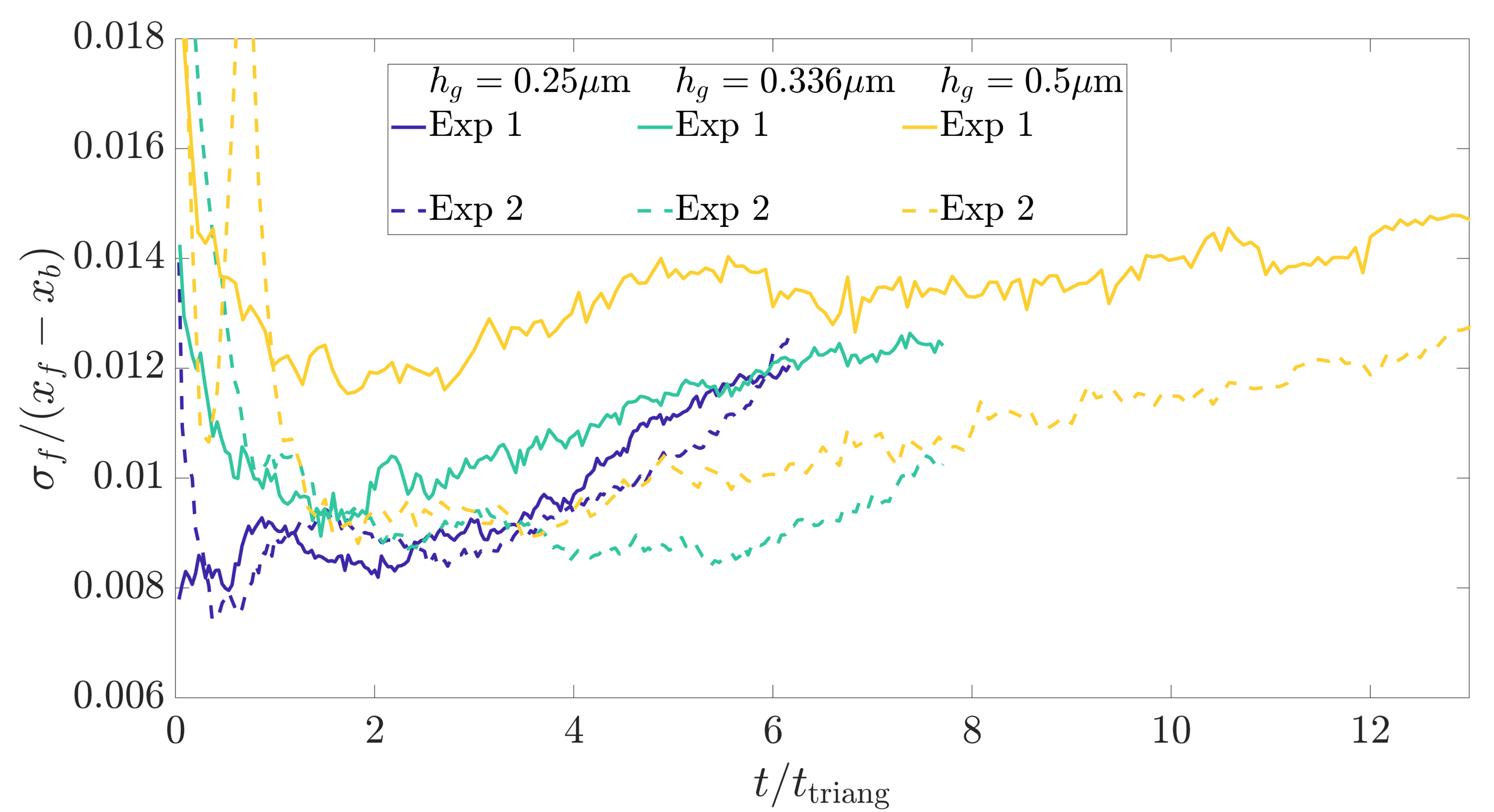}
\caption{\label{fig:sig_over_len} The scale of the roughness of the density
front $\sigma_{f}(t)=\text{std}\left(x_{f}\left(y;t\right)\right)$,
normalized by the length of the density profile $x_{f}(t)-x_{b}(t)$,
for two experiments at each of the three gravitational heights. Time
is scaled by a rough estimate of the time it takes for the density
profile to become (approximately) triangular, $t_{\text{triang}}=20$min
(see also \eqref{eq:t_triang}).}
\end{figure}

Importantly, the relative roughness of the density shock appears to
reach an approximately constant magnitude, with the front position
fluctuating over a range of about $100\mu$m at the final time. To
verify this, in Fig. \ref{fig:sig_over_len} we show the width of
the rough profile $\sigma_{f}(t)$, as measured from the standard
deviation of $x_{f}\left(y;t\right)$. We normalize the roughness
by the length of the density profile $x_{f}(t)-x_{b}(t)$ (predicted
to grow like $\sqrt{t}$ by the Burgers model \eqref{eq:x_f_Burgers}),
where henceforth we denote the mean position of the density front
with $x_{f}\left(t\right)=\av{x_{f}\left(y;t\right)}$. To estimate
the position of the back of the front $x_{b}(t)$, we fit a line to
the section of the density profile corresponding to the first fifth
of it's extent in the $x$ direction, and calculate $x_{b}$ as the
intercept of this line. This is reasonably robust except in the initial
stages of the experiment when the density profile is far from linear.

We see in Fig. \ref{fig:front_rough} that in all experiments the
roughness of the front remains quite small ($\sim1$\%) compared to
the length of the density profile, at all times after an initial transient
(predicted to be $t_{\mathrm{triang}}$ by the Burgers model, see
\eqref{eq:t_triang}) during which the triangular density profile
predicted by \eqref{eq:triangle_wave} is reached. Therefore, even
though the front roughness is large compared to the particle size
(as much as $50a$), it is still reasonable to approximate the density
profiles as effectively one-dimensional, $\phi(x,y,t)\approx\phi(x,t)$.
Thus, unless otherwise stated, we focus on analyzing the density $\phi(x,t)$
computed by averaging $\phi(x,y,t)$ along the $y$ direction.

\subsubsection{\label{subsec:SubBurgers}Sub-Burgers model}

In this section, we compare the predictions of the sub-Burgers model
\eqref{eq:subBurgers}, using the estimated collective sedimentation
velocity $v(\phi)$ shown in Fig. \ref{fig:v_phi}, to the Burgers
approximation \eqref{eq:Burgers}. Because particle tracking over
the length and time scales of interest was not possible in the experimental
studies of sedimenting Burgers fronts, we cannot do this comparison
for the experiments since we cannot accurately measure $\phi(x,t)$
in the experiments over the whole range of time. Instead, we use computer
simulations to evaluate the accuracy of the models \eqref{eq:subBurgers}
and \eqref{eq:Burgers}.

We do our best to make the initial conditions in the simulations mimic
the experiments by using the experimental scattered light intensity,
scaled using the total number of particles estimated by particle tracking,
as a proxy for $\phi(x,t)$. Since this is a particularly bad approximation
at the very high packing densities in the initial configuration in
the experiments, we skip an amount of time $T_{s}$ in the beginning
of each experiment so that the maximum packing density falls to around
$0.4$; the times $T_{s}$ are indicated in Table \ref{tab:sim_parameters}.
For the particle-based simulations, we generate initial conditions
from the experimentally measured $\phi(x,T_{s})$. For comparison,
we also numerically solve the sub-Burgers equation \eqref{eq:subBurgers}
using a high-resolution Godunov method\footnote{We thank Wenjun Zhao for implementing this method in one dimension
and improving the handling of limiting.} \cite{SemiLagrangianAdvection_2D} starting with $\phi(x,T_{s})$
as the initial condition.

\begin{figure}
\centering{}\includegraphics[width=1\textwidth]{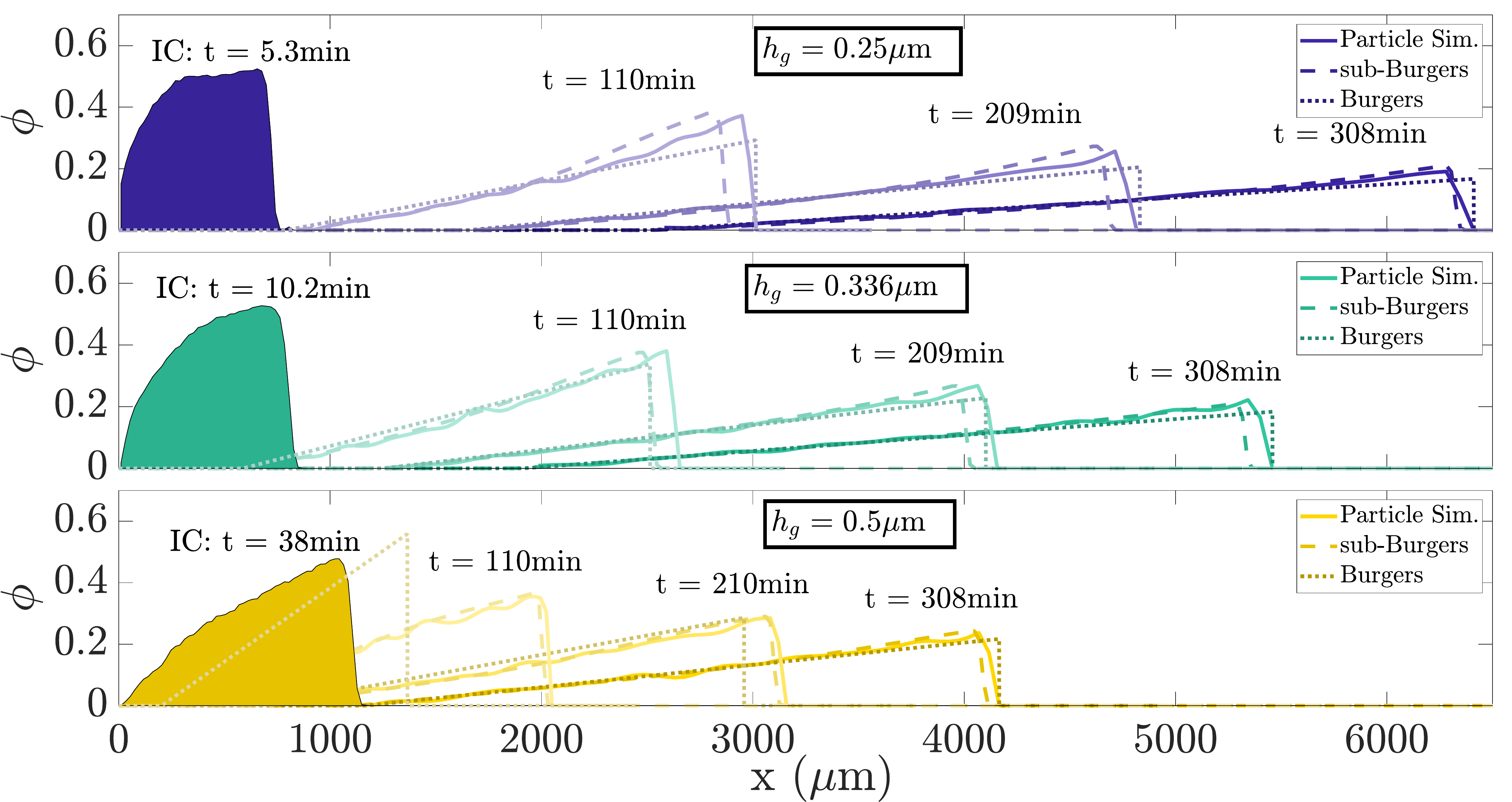}
\caption{\label{fig:phi_evo}Evolution of the density profile $\phi(x,t)$
for each gravitational height (Top: $h_{g}=0.25\mu$m, Middle: $h_{g}=0.33\mu$m,
Bottom: $h_{g}=0.5\mu$m). The filled density profile shows the initial
condition used in each case, taken from our experimental data after
an initial relaxation period (indicated in the figure). Each panel
shows the evolution of the density profile in a single particle simulation
(solid lines, results are reproducible in statistically independent
simulations), the sub-Burgers model \eqref{eq:subBurgers} (dashed
lines), and the Burgers approximation \eqref{eq:Burgers} (dotted
lines). Evolution of the density profiles in each case is shown using
decreasing transparency with increasing time (indicated in the figure).
An animated version of this figure is available in the Supplementary
Information.}
\end{figure}

The results of our computations are summarized in Fig. \ref{fig:phi_evo}.
We see that the initial condition quickly evolves into an approximately
triangular density profile, after which the sub-Burgers model \eqref{eq:subBurgers}
and the Burgers approximation \eqref{eq:Burgers} (see Section \ref{subsec:Burgers})
both agree quite well with the results from the particle simulations.
At early times, when the density profile is not yet triangular, the
sub-Burgers equation shows reasonable agreement with the particle
simulations, but the Burgers approximation does not, as expected.
This is particularly evident in the bottom panel of the figure for
$h_{g}=0.5\mu$m; similar behavior is seen at earlier times (not shown)
for the other values of $h_{g}$

\begin{figure}
\centering{}\includegraphics[width=1\textwidth]{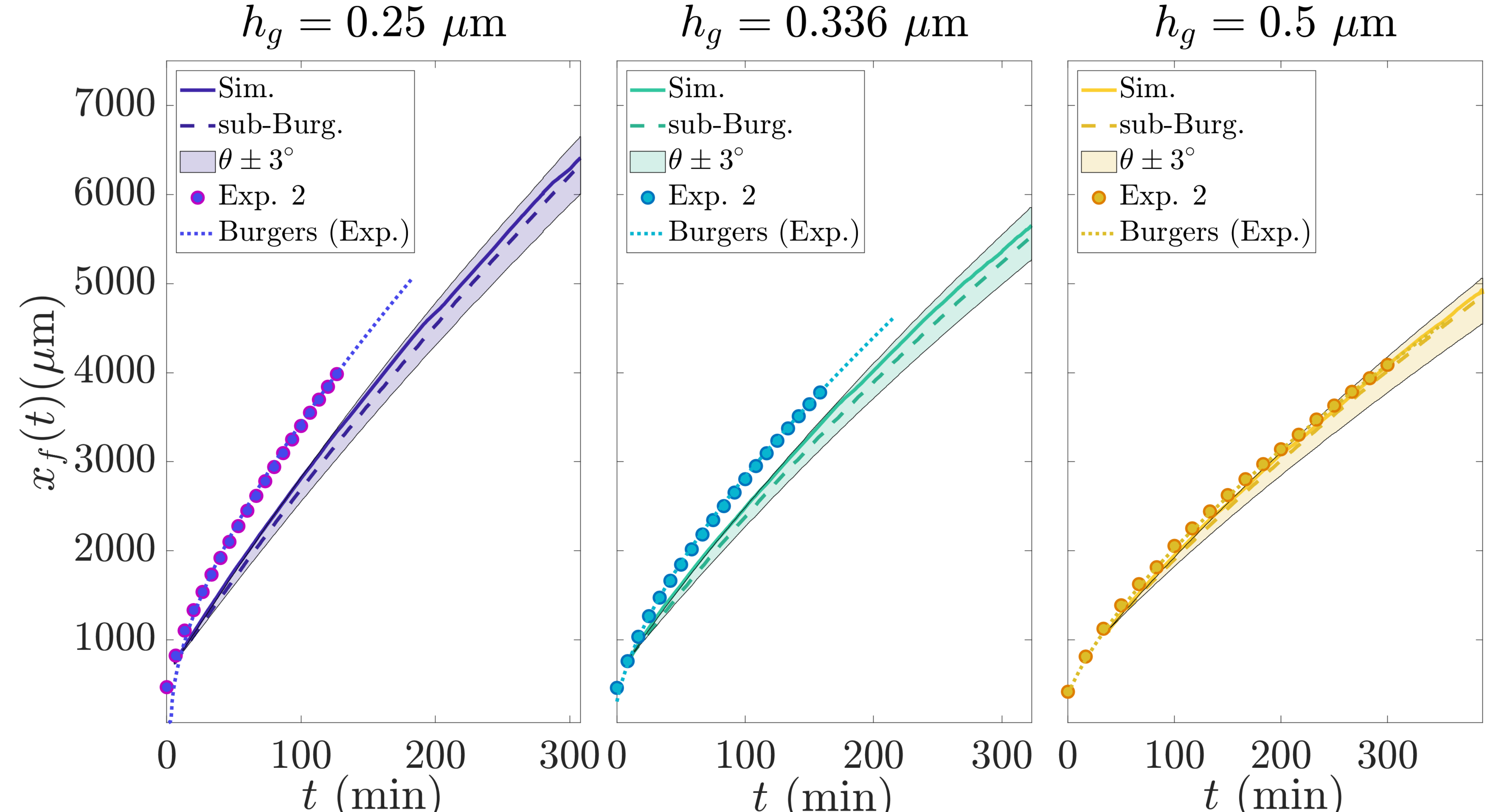}
\caption{\label{fig:sub_sim_exp}Mean position of the density front $x_{f}(t)$
for each of the three gravitational heights (increasing from left
to right). The experimental results (experiment \#2) are shown with
symbols (circles), and the Burgers fit to the experimental data (see
Section \ref{subsec:Burgers} for details) is shown with a dotted
line. We only show one of the experiments but the difference between
experiments is smaller than the symbol size. The results of particle
simulations are shown with a solid line along with corresponding shaded
regions bounded by the results obtained when the angle $\theta$ was
changed by $\pm3$ degrees, which is roughly the experimental uncertainty
in setting the microscope orientation. The predictions of the sub-Burgers
model \eqref{eq:subBurgers} are shown with a thick dashed line of
the same color as the particle results, while the Burgers approximation
to the simulations \eqref{eq:Burgers} is shown with a thin dashed
line of a different color (see legend).}
\end{figure}

We compare the predictions from our simulations to the experimental
measurements in Fig. \ref{fig:sub_sim_exp}. While we cannot measure
the density profile accurately in the first part of the experiments
due to the inability to track particles, $x_{f}(t)$ can be determined
accurately from the experimental images at all times. In Fig. \ref{fig:sub_sim_exp}
we compare predictions for the mean front position $x_{f}(t)$ between
the experiments, particle simulations, the sub-Burgers model \eqref{eq:subBurgers},
and the Burgers approximation \eqref{eq:Burgers}. The sub-Burgers
model is once again seen to agree with the particle simulations rather
well over the whole time interval, for all gravitational heights.
At the same time, we observe a moderate ($\sim10-15$\%) difference
between the particle simulations and experiments for the smallest
gravitational height $h_{g}=0.25\mu$m, with the agreement becoming
much better for the largest $h_{g}=0.5\mu$m. The systematic difference
persists even if we start simulations from snapshots of the experiments
at later times, when the density is lower and the scattered intensity
is a better proxy for the true density. This suggests that there is
a (yet unidentified) discrepancy in either the particle interactions
(e.g., electrostatics), or the hydrodynamics, between the simulations
and experiments \footnote{It should also be noted that a one-to-one comparison is difficult
to make because the width $L_{y}$ is much smaller in the simulations
to reduce the overall number of particles and thus control the computational
effort (e.g., $L_{y}\sim3.35$mm is the width of the viewing frame
in the experiments, while $L_{y}\sim100\mu$m in the simulations).}. At the largest $h_{g}$, the particles are furthest from each other
and the floor, and we expect the direct and hydrodynamic interactions
between particles to have less of an impact. We further compare experiments
and simulations quantitatively in Section \ref{subsec:Burgers}.

\subsubsection{\label{subsec:Burgers}Burgers model}

In order to compare our simulation results to the predictions of the
Burgers model \eqref{eq:Burgers}, we fit the mean position of the
front $x_{f}(t)$ and the back of the front $x_{b}(t)$ to the theoretical
prediction (\ref{eq:x_f_Burgers}), accounting also for the fact that
$v(\phi=0)$ is nonzero,
\begin{align}
x_{b}(t) & =v_{0}t+x_{0}\label{eq:x_b_t}\\
x_{f}(t)-x_{b}(t) & =\sqrt{2SM\left(t-t_{0}\right)},\label{eq:x_f_t}
\end{align}
where $v_{0}=v(\phi=0)$ is the numerically-estimated self velocity
of an isolated particle (see Table \ref{tab:Burgers}), $S=2\,dv\left(\phi=0\right)/d\phi$
is estimated from the slope of the rational fit to the simulation
data shown in Fig. \ref{fig:v_phi} (see Table \ref{tab:Burgers}),
and the total mass $M=N\,\pi a^{2}/L_{y}$. To account for the fact
that the initial condition is not a square wave, we obtain the time
shift $t_{0}$ and the position shift $x_{0}$ from $x_{f}\left(t_{f}\right)$
and $x_{b}\left(t_{f}\right)$ at the final time $t_{f}$. We remind
the reader that the Burgers equation predicts that any compactly supported
initial condition will asymptotically approach a triangle wave \eqref{eq:triangle_wave}
with a time and space shift that depends on the initial conditions.

\begin{table}
\caption{\label{tab:Burgers}Fitted or computed values of the parameters $S$
and $v_{0}$ in units of $\mu\text{m}/s$ in the Burger's model \eqref{eq:Burgers}.
For the particle simulations, the values are computed from the simulation
data shown in Fig. \ref{fig:v_phi} (see text). For two independent
experiments, the value of $S$ is obtained from the experimental data
shown in Fig. \ref{fig:sub_sim_exp} over the time period where $\phi(x,t)<0.4$
(see text). We also show estimates of $v_{0}$ obtained by fitting
the experimental $x_{b}(t)$ to \eqref{eq:x_b_t}.}

\begin{tabular}{|c|c|c|c|}
\hline 
$h_{g}$ ($\mu$m) & $0.25$ & $0.336$ & $0.5$\tabularnewline
\hline 
\hline 
Particle Sim. & $S=1.33$, $v_{0}=0.144$ & $S=1.11$, $v_{0}=0.108$ & $S=0.85$, $v_{0}=0.07$\tabularnewline
\hline 
Exp. \#1 & $S=1.61$, $v_{0}=0.154$ & $S=1.24$, $v_{0}=0.12$ & $S=0.732$, $v_{0}=0.0723$\tabularnewline
\hline 
Exp. \#2 & $S=1.67$, $v_{0}=0.167$ & $S=1.29$, $v_{0}=0.112$ & $S=0.766$, $v_{0}=0.0653$\tabularnewline
\hline 
\end{tabular}
\end{table}

In Table \ref{tab:Burgers} we compare the parameter $S$ and the
self velocity $v_{0}$ in the Burger's model \eqref{eq:Burgers} between
experiments and simulations. In the experiments, we can unambiguously
and accurately track the mean front position $x_{f}(t)$ over all
times, while the estimates for the end of the profile $x_{b}(t)$
are less reliable and more noisy, especially at early times. Nevertheless,
we estimate $v_{0}$ by fitting the experimental $x_{b}(t)$ to \eqref{eq:x_b_t}.
The results, shown in Table \ref{tab:Burgers}, demonstrate that $v_{0}$
is quite consistent between experiments and simulations. To obtain
$S,$ $t_{0},$ and $x_{0}$, we match the experimental data to Eqs.
(\ref{eq:x_b_t},\ref{eq:x_f_t}) at later times, when the triangular
density profile is well formed, by restricting the analysis to the
time interval $t_{i}<t<t_{f}$ after the maximum of the estimated
density is sufficiently low for the Burgers approximation to be apply,
$\max\phi(x,t_{i})<0.4$. We first determine $x_{0}$ from $x_{b}\left(t_{f}\right)$,
since the scattered intensity is the best proxy for the density at
the final time, using the self velocity $v_{0}$ estimated from simulations
(see Table \ref{tab:Burgers}) in order to avoid using the less reliable
$x_{b}\left(t_{i}\right)$. Then, we compute $S$ and $t_{0}$ from
$x_{f}\left(t_{i}\right)$ and $x_{f}\left(t_{f}\right)$. The resulting
Burgers prediction \eqref{eq:x_f_t} with the estimated parameters
is shown in Fig. \ref{fig:sub_sim_exp}, and is in excellent agreement
with the experimental data over nearly the whole time interval. The
experimental estimates of the parameter $S$ in Table \ref{tab:Burgers}
show the same qualitative trends with gravitational height, but an
imperfect quantitative agreement, between simulations and experiments,
as already seen in Fig. \ref{fig:sub_sim_exp}.

\begin{figure}
\centering{}\includegraphics[width=1\textwidth]{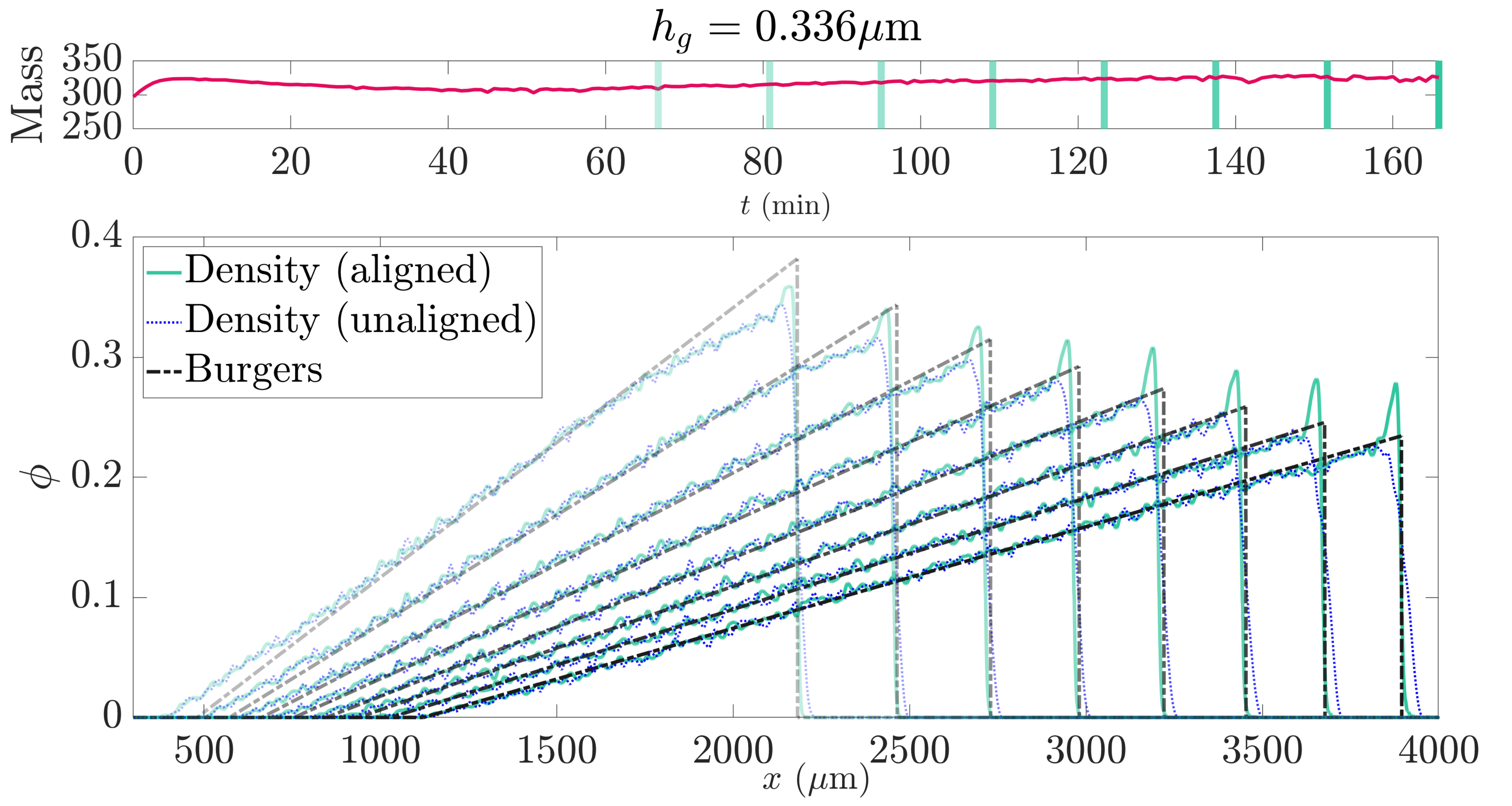}
\caption{\label{fig:Exp_Burg_Fit}The top panel shows the total mass of the
density profile $M={\displaystyle \int\phi(x,t)dx}$ for experiment
$\#2$ at $h_{g}=0.336$, as measured by scattered intensity re-scaled
to be in units of particle density based on the estimated total number
of particles $N_{\text{exp}}$. The mass remains constant to within
5\% after about $t_{i}=70.8$min, so we compare the experiment data
to the Burgers model for $t>t_{i}$. The bottom panel compares snapshots
in time of the Burgers model (\ref{eq:triangle_wave}) along with
the experimental measurements of the density profile. The curves have
been colored to increase in opacity as time increases, and the times
of each snapshot are indicated by vertical bars in the top panel using
a corresponding transparency. The experimental measurements of the
density profile have been averaged along the $y$ direction either
without (labeled ``unaligned'') or with alignment (labeled ``aligned'')
of the front $x_{f}\left(y;t\right)$ (see text). An animated version
of this figure is available in the Supplementary Information.}
\end{figure}

Finally, in Fig. \ref{fig:Exp_Burg_Fit} we compare the Burgers prediction
(\ref{eq:triangle_wave},\ref{eq:x_f_Burgers}) for the density profile
to experimental data. For this comparison, we focus on experiment
$\#2$ at $h_{g}=0.336$, for which the total scattered intensity
(proportional to the total mass $M$) was constant for $t_{f}\geq t>t_{i}=70.8$min
to within $5$\%, suggesting that the scattered intensity was a good
proxy for $\phi(x,t>t_{i})$. The scattered intensity has a small
(at most couple of percent of max intensity) constant background intensity
(varying slowly over time) both in front of and behind the support
of $\phi$. To compute the mass of the density profiles we offset
$\phi(x,t)$ to remove the estimated background intensity, and match
$M=M_{\text{exp}}$ at the final time $t_{f}$. We estimate the position
of the back $x_{b}\left(t_{f}\right)$ by first estimating the position
of the front $x_{f}\left(t_{f}\right)$ and then finding the value
of $x_{b}\left(t_{f}\right)$ that minimizes the $L_{2}$ error between
$\phi(x,t_{f})$ and a triangle of area $M$ extending from $x_{f}\left(t_{f}\right)$
to $x_{b}\left(t_{f}\right)$. Using the experimental data for \footnote{We note that the experimental $x_{b}(t>t_{i})$ was in excellent agreement
with the theoretical prediction \eqref{eq:x_b_t} and the value of
$v_{0}$ extracted from the simulations (see Table \ref{tab:Burgers}).} $x_{b}\left(t_{f}\right)$, $x_{f}\left(t_{f}\right),$ and $x_{f}\left(t_{i}\right)$
and Eqs. (\ref{eq:x_b_t},\ref{eq:x_f_t}), we estimate $t_{0}=6.8$min
and $S=1.24$ (compare to data in Table \ref{tab:Burgers}). In Fig.
\ref{fig:Exp_Burg_Fit} we compare the theoretical prediction for
the triangular density profile (\ref{eq:triangle_wave}) with the
extracted parameters to the experimental density profile $\phi(x,t)$
obtained by averaging the experimentally measured $\phi(x,y,t)$ along
$y$, and subtracting the offset due to the nonzero background intensity.
We also compute the average density along $y$ by first shifting the
profile $\phi\left(x,y,t\right)$ for each $y$ in the $x$ direction
by $x_{f}\left(y;t\right)-\av{x_{f}\left(y;t\right)}$. By construction,
this shifting preserves the mean and therefore the aligned density
profile has the same mean front position $x_{f}(t)$. The alignment
does however allow us to examine the variation in density along the
$x$ direction at scales comparable to the particle size, which is
not possible with the un-aligned profiles because of the smearing
caused by the roughness of the front.

We observe in Fig. \ref{fig:Exp_Burg_Fit} that both the aligned and
un-aligned density profiles exhibit the expected triangular shape,
and are in reasonable agreement with the Burgers prediction. The aligned
data shows a sharp density front, but also exhibits a small peak at
the front. Numerical simulations suggest that right at the front,
the particles get lifted above the floor to heights larger than $h_{g}$,
and can therefore reach higher in-plane packing densities. The height
of the particles at the front depends on the balance between the self-induced
motion due to gravity (pulling them toward the wall) and the upward
flow away from the wall, induced by the force acting on the particles
behind the front. This effect is more pronounced for larger incline
angles $\theta$, and for sufficiently large $\theta$ we find that
the particles do not remain in a monolayer, especially near the density
maximum at the front.

\section{\label{sec:Conclusions}Conclusions}

We examined the driven evolution of the planar density profile of
a colloidal monolayer sedimenting down an inclined plane, using experiments
and simulations. We found that, starting from an approximately constant
density in a stripe of finite width, the monolayer develops an inhomogeneous
triangular density profile with a shock front at the front edge. At
the same time, the front becomes rough, but the \emph{relative }roughness
saturates to an approximately constant amplitude and remains stable.
We found that a simple one-dimensional sub-Burgers equation predicts
the shape of the density profile well over all times, and a Burgers
approximation is accurate at later times when the density is sufficiently
low. These simple models only require as input the collective sedimentation
velocity of a uniform suspension at a given density, which can be
obtained easily in either experiments or simulations. The Burgers
equation can be solved analytically and only requires a single input,
the slope of the collective velocity as a function of density. We
found a modest but systematic difference in the Burgers parameters
estimated from fitting experimental data and from simulations, perhaps
attributable to unaccounted interactions between the particles and
the particles and the floor.

There are many avenues for improving the simple Burgers model we focused
on in this work, to try to improve the match between theory, simulations,
and experiments. For example, since the roughness of the density front
is large compared to the particle size, a natural step would be to
consider a two-dimensional conservation law instead of a one-dimensional
one. However, the two-dimensional Burgers equation is stable to transverse
perturbations \cite{Burgers2D_ShockStability} and would not develop
the roughness we observe.

For colloidal microrollers, the local mean-field approximation \eqref{eq:subBurgers}
is not appropriate. Assuming the colloids lie approximately in a plane
at height $h$ above the bottom floor, it was shown in \cite{NonlocalShocks_Rollers}
that one should instead use a \emph{nonlocal} conservation equation
\begin{eqnarray}
\frac{\partial\rho}{\partial t} & \approx & -\frac{\partial\left[\rho\left(K_{\text{roll}}^{(h)}*\rho\right)\right]}{\partial x},\label{eq:nonlocal_shock}
\end{eqnarray}
where star denotes convolution with a translationally-invariant kernel
\[
K_{\text{roll}}^{(h)}(x)\sim hx^{2}/\left(x^{2}+4h^{2}\right)^{2}
\]
that describes the flow field created by a unit rotlet above a no-slip
plane \cite{blake1971note,OseenBlake_FMM}. The kernel $K_{\text{roll}}^{(h)}(x)$
is zero at the origin, which shows that the collective motion arises
due to \emph{non-local }hydrodynamic interactions of particles at
distances $O(h)$ rather than neighboring particles. However, the
Blake solution for a Stokeslet above a no-slip plane \cite{blake1971note,OseenBlake_FMM}
can be used to show that the corresponding kernel for sedimentation
\[
K_{\text{sed}}^{(h)}(x)\sim4h^{2}\frac{3x^{2}+4h^{2}}{\left(x^{2}+4h^{2}\right)^{2}}+\ln\left(1+\frac{4h^{2}}{x^{2}}\right)
\]
is singular at and peaked around the origin, which suggests that collective
motion arises primarily due to \emph{local }hydrodynamic interactions
among neighboring particles. It is therefore reasonable to expect
that the local approximation \eqref{eq:subBurgers} is reasonable
for sedimentation \footnote{In fact, our attempts to solve the nonlocal equation \eqref{eq:nonlocal_shock}
with the kernel $K_{\text{sed}}^{(h)}(x)$ (or the corresponding two-dimensional
generalization) numerically indicate that the nonlocal equation develops
singular solutions where all the particles clump together at the shock
front reaching unphysically high densities in finite time. This unphysical
behavior arises because steric repulsion is not accounted for in the
simple model \eqref{eq:nonlocal_shock}; it is not straightforward
to add particle repulsion in continuum models.}.

Since we know that the hydrodynamic interactions driving the collective
dynamics are non-local and the front is rough, it is natural to consider
(two-dimensional) non-local conservation laws. However, we find that
steric repulsion has to be accounted for in such models because otherwise
the density develops unphysically-large peaks near the front. This
is not straightforward to do and would lead to nonlinear nonlocal
conservation laws which could not be solved numerically using standard
methods (in fact, performing the particle simulations is likely just
as fast and simpler!), nor would they lead to the physical insight
that the simple Burgers model does. In the end, the success of the
Burgers equation in describing the collective density dynamics is
unexpected but in large part owed to its simplicity and to the universal
nature of Burgers shocks at long times \cite{HyperbolicLaws_Lax}.
\begin{acknowledgments}
We thank Wenjun Zhao for sharing with us her code for a high-resolution
advection solver in one dimension. We also thank Jonathan Goodman
for helpful discussions regarding the Burgers equation. This work
was supported primarily by the MRSEC Program under award DMR-1420073.
Additional funding was provided by the National Science Foundation
under award number CBET-1706562. B.S. and A.D. were supported by the
National Science Foundation via the Research Training Group in Modeling
and Simulation under award RTG/DMS-1646339. B.S. and A.D. also thank
the NVIDIA Academic Partnership program for providing GPU hardware
for performing the simulations reported here. P. Chaikin was partially
supported by NASA under Grant Number NNX13AR67G.
\end{acknowledgments}

\bibliographystyle{unsrt}
\bibliography{References,RollersShocks}

\end{document}